\documentclass[preprint,tightenlines,aps,prd,showpacs]{revtex4}
\usepackage{graphicx}
\usepackage{amsmath}
\usepackage{bm}

\begin{document}

\title{Inclusive Production of the $X(3872)$}
%%%%%%%%%%%%%%%%%%%%%%%%%%%%%%%%%%%%%%%%%%%%%%%%%%%%%%%%%%%%%%%%%%%%%%%%%%%%%%

\author{Eric Braaten}
%\thanks{}
\affiliation{
Physics Department, Ohio State University, 
Columbus, Ohio 43210, USA}

\date{\today}
%%%%%%%%%%%%%%%%%%%%%%%%%%%%%%%%%%%%%%%%%%%%%%%%%%%%%%%%%%%%%%%%%%%%%%%%%%%%%%
\begin{abstract}
If the $X(3872)$ is a loosely-bound 
$D^{*0} \bar D^0$/$D^0 \bar D^{*0}$ molecule, 
its inclusive production rate can be described by the 
NRQCD factorization formalism that applies to inclusive 
quarkonium production.  We argue that if the molecule has 
quantum numbers $J^{PC} = 1^{++}$, the most important term in the 
factorization formula should be the color-octet $^3S_1$ term.
This is also one of the two most important terms in the 
factorization formulas for $\chi_{cJ}$.
Since the color-octet $^3S_1$ term dominates $\chi_{cJ}$ 
production for many processes, 
the ratio of the inclusive direct production rates for 
$X$ and $\chi_{cJ}$ should be roughly the same for these processes. 
The assumption that the ratio of the production 
rates for $X$ and $\chi_{cJ}$ is the same for all processes
is used to estimate the inclusive 
production rate of $X$ in $B$ meson decays, $Z^0$ decays, 
and in $p \bar p$ collisions.
\end{abstract}

\pacs{12.38.-t, 12.38.Bx, 13.20.Gd, 14.40.Gx}
% 12.38.-t   Quantum chromodynamics
% 12.38.Bx   Perturbative calculations
% 13.20.Gd  Decays of J/psi, Upsilon, and other quarkonia
% 14.40.Gx   Mesons with S=C=B=0, mass > 2.5 GeV (including quarkonia)

%\keywords{}

\maketitle


The $X(3872)$ is a mysterious charmonium-like state discovered by 
the Belle collaboration in 2003 \cite{Choi:2003ue}.  
It was discovered through the exclusive decay $B^+ \to X K^+$ 
followed by the decay $X \to J/\psi \; \pi^+ \pi^-$.  
The discovery was confirmed by the CDF collaboration 
\cite{Acosta:2003zx}, which observed the inclusive production 
of the $X$ meson in $p \bar p$ collisions.  
The $X$ has also been observed in the discovery mode 
by the BaBar collaboration \cite{Aubert:2004ns} and in 
$p \bar p$ collisions by the D0 collaboration \cite{Abazov:2004kp}.

The interpretation of the $X$ meson remains a mystery
\cite{Tornqvist:2004qy,Close:2003sg,Pakvasa:2003ea,%
Braaten:2003he,Voloshin:2003nt,Wong:2003xk,Barnes:2003vb,%
Swanson:2003tb,Eichten:2004uh,Kim:2004cz}.  
The decay mode $J /\psi \; \pi^+ \pi^-$ suggests that it is 
a charmonium state, but its decay properties seem to be incompatible 
with its identification with any charmonium state 
\cite{Barnes:2003vb,Eichten:2004uh,Olsen:2004fp}.  
The Belle collaboration found that the width of $X$ is small
compared to that of other charmonium states above the 
$D \bar D$ threshold:  
$\Gamma_X < 2$ MeV at the 90\% confidence level \cite{Choi:2003ue}.
In addition to the discovery decay mode 
$J /\psi \; \pi^+ \pi^-$,
the $X(3872)$ has been observed in the decay modes 
$J /\psi \; \pi^+ \pi^- \pi^0$ and $J /\psi \; \gamma$ \cite{Abe:2005iy}.
Upper limits have been placed on the branching fractions for
several other decay modes of the $X$, including
$D^0 \bar D^0$, $D^+ D^-$, $D^0 \bar D^0 \pi^0$ \cite{Abe:2003zv},
$\chi_{c1} \gamma$, $\chi_{c2} \gamma$, 
$J/\psi \, \pi^0 \pi^0$ \cite{Abe:2004sd},
and $J/\psi \, \eta$ \cite{Aubert:2004fc}.

Another possibility is that the $X$ meson is a 
$D^{*0} \bar D^0$/$D^0 \bar D^{*0}$ molecule 
\cite{Tornqvist:2004qy,Voloshin:2003nt,Braaten:2003he,Swanson:2003tb}.  
This possibility is motivated by the fact that its mass is 
extremely close to the $D^{*0} \bar D^0$ threshold.
The mass obtained by combining the measurements 
of the four experiments is $M_X = 3871.9 \pm 0.5$ MeV \cite{Olsen:2004fp}.
Subtracting the $D^{*0} \bar D^0$ threshold energy, we obtain
$M_X - M_{D^0} - M_{D^{*0}} = +0.6 \pm 1.1$ MeV.  
For this molecule to have a substantial 
branching fraction into $J /\psi \; \pi^+ \pi^-$, it must be a bound 
state, which requires that its binding energy 
$E_X=M_{D^0} + M_{D^{*0}}- M_X$ be positive.
(Note that the central value of $E_X$ from the mass
measurement is negative.) 
If the binding of the $D^{*0} \bar D^0$ molecule arises 
from pion exchange, the most likely channel is S-wave with 
quantum numbers $J^{PC} = 1^{++}$ \cite{Tornqvist:2004qy}.  
The observed binding energy $E_X$ 
is small compared to the natural energy scale associated with 
pion-exchange interactions:  $E_X \ll m^2_\pi / 2 \mu \approx 10$ MeV, 
where $\mu$ is the reduced mass of the $D^{*0}$ and $\bar D^0$. 
The small binding energy implies that
the $D^{*0} \bar D^0$ scattering length $a$ is large compared to 
the natural length scale $1/m_\pi \approx 1.4$ fm.  As a consequence of the 
large scattering length, the molecule has universal properties 
that are determined by $a$ and insensitive to the 
shorter length scales of QCD \cite{Braaten:2003he}.  
The simplest example is the binding energy itself:  
$E_X =1/(2 \mu a^2)$.  
Another example is the rate and momentum spectrum for the decays 
$X \to D^0 \bar D^0 \pi^0$ and $X \to D^0 \bar D^0 \gamma$ 
\cite{Voloshin:2003nt}.  A third example is the rate for the decay 
$\Upsilon (4S) \to X h^+ h^-$, where $h^+$ and $h^-$ 
are light charged hadrons \cite{Braaten:2004rn}.  
Another universal feature associated with the large 
scattering length is an enhancement in the $D^{*0} \bar D^0$ 
invariant mass distribution near the $D^{*0} \bar D^0$ threshold 
in the decay $B^+ \to D^{*0} \bar D^0 K^+$ \cite{Braaten:2004fk}.

A challenge to any interpretation of the $X(3872)$ is to understand its 
production rate.  The Belle amd Babar collaborations have measured 
the product of the branching fractions for the two decays associated with 
its discovery \cite{Choi:2003ue,Aubert:2004ns}:
\begin{eqnarray}
{\rm Br} [ B^+ \to XK^+ ] \; {\rm Br} [ X \to J/\psi \; \pi^+ \pi^-] 
= (1.3 \pm 0.3) \times 10^{-5}. 
\label{Belle-prod}
\end{eqnarray}
The only other information about the production of $X$ that is available is 
that its inclusive production has been observed in $p \bar p$ collisions.

In this paper, we will discuss the inclusive production of the $X(3872)$.  
We will point out that if the $X$ meson is a loosely-bound S-wave 
$D^{*0} \bar D^0$/$D^0 \Bar D^{*0}$ 
molecule, its inclusive production rate can be analyzed using the NRQCD 
factorization formalism \cite{Bodwin:1994jh}.  
We will argue that if its quantum numbers are 
$1^{++}$, the most important term in the NRQCD factorization formula 
is the color-octet $^3S_1$ term. This is also one of the two most important 
terms in the NRQCD factorization formulas for $\chi_{cJ}$.  
Thus for all processes in which $\chi_{cJ}$ production is dominated by
the color-octet $^3S_1$ term, the ratio of the inclusive production rates 
for $X$ and $\chi_{cJ}$ should be roughly the same.  
The assumption that the ratio of the inclusive production rates 
for $X$ and $\chi_{cJ}$ should be the same 
will be used to estimate the inclusive 
production rate of $X$ in $B$ meson decays, $Z^0$ decays, 
and in $p \bar p$ collisions.

The production of the $X$ meson involves many length scales.  
The creation of the $c \bar c$ pair with small relative momentum 
requires a hard-scattering process with a short-distance 
length scale $1/m_c$.  The evolution of the $c \bar c$ into a 
color-singlet hadronic system, either a charmonium or a 
pair of charmed mesons, occurs over a longer length scale 
$1 / m_c v$ or $1 / m_c v^2$.  
The evolution of the charmed mesons into one of the four 
possibilities $D \bar D$, $D^*\bar D$, $D \bar D^*$, 
and $D^* \bar D^*$ occurs over an even longer length 
scale $1 /m_\pi$.  Finally the binding of $D^{*0} \bar D^0$ into the 
molecular state $X$ occurs over the very long length scale $a$.  
We will use the NRQCD factorization formalism to exploit the 
separation between the short-distance scale $m_c$ and all the 
longer distance scales of order $1/m_cv$ and larger.

The NRQCD factorization formalism \cite{Bodwin:1994jh} is used 
to describe the inclusive production of charmonium.  
This formalism exploits the separation of scales between 
the hard-scattering process that creates the $c \bar c$ pair 
and the hadronization process by which the $c \bar c$ pair evolves 
into a color-singlet charmonium.  The NRQCD factorization formula 
expresses the inclusive cross section for producing the 
charmonium state $H$ as a sum of products of $c \bar c$ cross sections 
and NRQCD matrix elements:
\begin{eqnarray}
\sigma [ H + {\rm anything}] = \sum_n \hat \sigma [ c \bar c_n + {\rm anything}] 
\; \langle {\cal O}_n^H \rangle .
\label{ff:H}
\end{eqnarray}
The $c \bar c$ cross section $\hat \sigma$ is the inclusive cross section 
for producing a $c \bar c$ pair in the color and angular-momentum 
channel labeled by $n$.  It involves momentum 
scales of order $m_c$ and larger.  
The corresponding NRQCD matrix element $\langle {\cal O}_n^H \rangle$ 
is the vacuum expectation value of an operator that creates 
a $c \bar c$ pair at a point in the channel specified by $n$, 
projects onto states that in the asymptotic future
include the charmonium $H$, and then annihilates the $c \bar c$ pair 
at the creation point in the channel specified by $n$.  
The NRQCD matrix elements involve only  
momentum scales of order $m_c v$ or smaller, where $v$ is the typical 
relative velocity of the charm quark in the charmonium.  
The sum in (\ref{ff:H}) includes infinitely many terms, 
so it is useful only if the sum over channels can be truncated.
The truncation can be justified by the 
velocity-scaling rules for quarkonium, which specify how each 
of the NRQCD matrix elements scale with $v$.
For P-wave charmonium states, there are two NRQCD matrix elements 
at leading order in $v$:  a color-singlet P-wave matrix element 
and a color-octet S-wave matrix element \cite{Bodwin:1992ye}.  
If the NRQCD expansion is truncated at this order, 
the inclusive cross section for $\chi_{cJ}$ can be written
\begin{eqnarray}
\sigma [ \chi_{cJ} + {\rm anything} ] &\simeq& 
\hat \sigma [ c \bar c_1 ({}^3P_J) + {\rm anything} ]  \;
	\langle {\cal O}_1^{\chi_{cJ}} ({}^3P_J) \rangle 
\nonumber
\\
&+& \hat \sigma [ c \bar c_8 ({}^3S_1) + {\rm anything}]  \;
	\langle {\cal O}_8^{\chi_{cJ}} ({}^3S_1) \rangle .
\label{fft:chi1}
\end{eqnarray}
The subscript on $c \bar c$ and on the operator ${\cal O}$ 
indicates the color channel and the argument in parentheses 
indicates the angular-momentum channel.
The relative sizes of the two terms in Eq.~(\ref{fft:chi1})
depends on the NRQCD renormalization scale \cite{Bodwin:1992ye}.
A change in the NRQCD renormalization scale for the operator
${\cal O}_1 ({}^3P_J)$ can be compensated by adding the operator 
${\cal O}_8({}^3S_1)$ with a coefficient of order $\alpha_s$.
For processes in which a color-octet $c \bar c$ pair
can be created at leading order in $\alpha_s$,
the color-octet $^3S_1$ term in the cross section is likely 
to dominate for a suitable choice of the renormalization scale.
Heavy quark spin symmetry implies that the matrix elements in 
Eq.~(\ref{fft:chi1}) are proportional to $2J+1$ \cite{Bodwin:1994jh}.
Thus the factorization formula can be written
\begin{eqnarray}
\sigma [ \chi_{cJ} + {\rm anything} ] &\simeq& 
(2J+1) \hat \sigma [ c \bar c_1 ({}^3P_J) + {\rm anything} ]  \;
	\langle {\cal O}_1^{\chi_{c0}} ({}^3P_0) \rangle 
\nonumber
\\
&+& (2J+1) \hat \sigma [ c \bar c_8 ({}^3S_1) + {\rm anything}]  \;
	\langle {\cal O}_8^{\chi_{c0}} ({}^3S_1) \rangle .
\label{fft:chi2}
\end{eqnarray}
The only dependence on $J$ in the coefficient of 
$\langle {\cal O}_8^{\chi_{c0}} ({}^3S_1) \rangle$
is the explicit factor of $2J+1$.
Thus processes that are dominated by the color-octet ${}^3S_1$
term can be identified experimentally by the cross sections for 
$\chi_{c1}$ and $\chi_{c2}$ being in the proportions 3:5.

The NRQCD factorization formalism applies equally well 
to a pair of charm mesons, such as $D^{*0}$ and $\bar D^0$,
that are produced with a relative momentum of order 
$m_c v$ or smaller.  For example, the inclusive cross section 
for producing a $D^{*0} \bar D^0$ pair with kinetic energy
in its rest frame
less than $E_{\rm max}$, where $E_{\rm max} \ll m_c$, can be written
\begin{eqnarray}
\sigma [ D^* \bar D(E_{\rm max}) + {\rm anything} ]
= \sum_n \hat 
\sigma [ c \bar c_n + {\rm anything} ]  \;
\langle {\cal O}_n^{D^* \bar D(E_{\rm max})} \rangle.
\label{ff:DbarDstar}
\end{eqnarray}
We have used $D^* \bar D(E_{\rm max})$ as a short-hand for 
$D^{*0} \bar D^0$ states with kinetic energy less than $E_{\rm max}$.
The $c \bar c$ cross sections $\hat \sigma$  are the same as in
(\ref{ff:H}).  In the matrix elements, the NRQCD operators 
include a projection onto states that in the asymptotic
future include a $D^{*0} \bar D^0$ pair with kinetic energy
less than $E_{\rm max}$.
If the $X$ meson is a loosely-bound $D^{*0} \bar D^0$/$D^0 \bar D^{*0}$ 
molecule, the NRQCD factorization formula 
can also be applied to its inclusive production rate.  
The inclusive cross section for producing $X$ can be written
\begin{eqnarray}
\sigma [ X + {\rm anything} ]
= \sum_n \hat 
\sigma [ c \bar c_n + {\rm anything} ]  \;
\langle {\cal O}_n^X \rangle.
\label{ff:X}
\end{eqnarray}
In the matrix elements, the NRQCD operators 
include a projection onto states that in the asymptotic
future include an $X(3872)$.

The NRQCD factorization formulas in Eqs.~(\ref{ff:DbarDstar})
and (\ref{ff:X}) are useful only if the expansions 
can be truncated.  Unlike the case of quarkonium, 
there are no velocity scaling rules governing the NRQCD 
matrix elements for $D^{*0} \bar D^0$ or $X$ that can be used to 
justify the truncations.  The velocity-scaling rules would be applicable 
if the $c \bar c$ pair evolved into a heavy hadron $H$ with
constituents $c$ and $\bar c$ only through the emission of real 
or virtual gluons from the $c$ or $\bar c$. 
Velocity-scaling rules do not apply if the evolution involves
the emission of real or virtual gluons from light quarks 
or antiquarks in the hadron $H$.  Thus the presence of $\bar u$ 
and $u$ as valence quarks in the $D^{*0}$ and $\bar D^0$
guarantees that the velocity scaling rules are inapplicable.

Another class of $c \bar c$ systems for which the velocity scaling rules 
are inapplicable is charmonium hybrid states, whose valence content 
is $c \bar c g$ \cite{Chiladze:1998ti}.
Chiladze, Falk and Petrov pointed out that the NRQCD factorization 
formula could nevertheless be applied to the inclusive production of 
charmonium hybrid states.  We can use similar arguments to justify
the NRQCD factorization formula for the inclusive production of $X$.
The NRQCD matrix elements in the NRQCD factorization formula 
for inclusive $X$ production in Eq.~(\ref{ff:X})
are proportional to the probabilities 
for the evolution of the $c \bar c$ pair into a color-singlet 
$c \bar c u \bar u$ system, its subsequent evolution into an S-wave 
$D^{*0} \bar D^0$ or $D^0 \bar D^{*0}$ system, and finally 
the binding of the charm mesons into the molecular state $X$.  
Since the $X$ is weakly bound,
the charm mesons are almost at rest. 
The momentum scale of the $c$ and $\bar c$ in the $X$ is therefore
that of the $c$ and $\bar c$ in the charm mesons, 
which is of order $\Lambda_{\rm QCD}$.
This is small compared to the momentum scale 
of the hard-scattering process that creates the $c \bar c$ pair,
which is of order $m_c$ or larger.
This guarantees that the hard scattering amplitude can be expanded
in powers of the relative momentum of the $c \bar c$.
The convergence of this expansion must come from a suppression 
of the NRQCD matrix elements of higher dimension operators.  
The convergence suggests a hierarchy of NRQCD matrix elements 
according to the dimension of the NRQCD operators.
In particular, matrix elements associated with $c \bar c$ channels 
with higher orbital angular momentum quantum numbers should be 
suppressed.  This motivates a truncation to a few 
of the lowest dimension operators in each of the four color/spin 
channels:  color-singlet spin-singlet, color-singlet spin-triplet,
color-octet spin-singlet, and color-triplet spin-triplet.
The most extreme truncation of the NRQCD factorization formula
would be to include only the four S-wave matrix elements:
$\langle {\cal O}_1^X ({}^1S_0) \rangle$,
$\langle {\cal O}_1^X ({}^3S_1) \rangle$,
$\langle {\cal O}_8^X ({}^1S_0) \rangle$, 
and $\langle {\cal O}_8^X ({}^3S_1) \rangle$.

Voloshin has pointed out that if the $X(3872)$ is a 
$D^{*0} \bar D^0$ molecule with $J^{PC} = 1^{++}$, the $c \bar c$ pair 
in the $X$ is necessarily in a spin-triplet state \cite{Voloshin:2004mh}.
Heavy-quark spin symmetry implies that
$X(3872)$ is produced predominantly from $c \bar c$ 
pairs that are created in spin-triplet states.
Thus the spin-singlet matrix elements for the production of 
$X(3872)$ are suppressed.
If the  NRQCD factorization formula is truncated to include 
only the S-wave matrix elements, the leading terms are
\begin{eqnarray}
\sigma [ X + {\rm anything} ] &\approx& 
\hat \sigma [ c \bar c_1 ({}^3S_1 ) + {\rm anything} ]  \;
	\langle {\cal O}_1^X ({}^3S_1) \rangle 
\nonumber
\\
&+& \hat \sigma [ c \bar c_8 ({}^3S_1 ) + {\rm anything} ]  \;
	\langle {\cal O}_8^X ({}^3S_1) \rangle .
\label{fft:X2}
\end{eqnarray}

We now argue that the color-octet term in Eq.~(\ref{fft:X2}) 
will often dominate.
First we note that since the constituent charm mesons in the $X$ 
are color-singlets, the probabilities for the $c \bar c$ pair
in the $X$ to be in a color-singlet and color-octet state 
are $1/9$ and $8/9$, respectively.
The naive expectation is that the probabilities for forming 
an $X$ from the color-singlet $c \bar c$ pair 
and from each of the 8 color-octet $c \bar c$ pairs should be equal.
Thus unless the color-singlet parton cross section $\hat \sigma$
is larger than the color-octet parton cross section
by a factor large enough to compensate for the color factor of 8,
we would expect the color-octet term in Eq.~(\ref{fft:X2})
to dominate.  This reduces the NRQCD factorization formula to
a single term: 
\begin{eqnarray}
\sigma [ X + {\rm anything} ] &\approx& 
\hat \sigma [ c \bar c_8 ({}^3S_1 ) + {\rm anything} ]  \;
	\langle {\cal O}_8^X ({}^3S_1) \rangle .
\label{fft:X1}
\end{eqnarray}

We have argued that the color-octet ${}^3S_1$ term should be the most 
important term in the NRQCD factorization formula for the inclusive 
production of $X$, in which case it reduces to Eq.~(\ref{fft:X1}).
The color-octet ${}^3S_1$ term is also one of the two leading terms
in the NRQCD factorization formula for $\chi _{cJ}$,
which is given in Eq.~(\ref{fft:chi1}).  For many production processes, 
the color-octet ${}^3S_1$ term in the NRQCD factorization formula 
for $\chi _{cJ}$ will dominate for a suitable choice of the 
NRQCD renormalization scale.  For these processes, the ratio of the 
inclusive cross sections for $X$ and $\chi _{cJ}$ will be the ratio 
of the color-octet ${}^3S_1$ matrix elements:
\begin{eqnarray}
{\sigma [ X + {\rm anything}] 
	\over \sigma [\chi_{cJ} + {\rm anything}]} 
\approx {\langle {\cal O}_8^X ({}^3S_1) \rangle 
	\over (2J+1) \langle {\cal O}_8^{\chi_{c0}} ({}^3S_1) \rangle}.
\label{assume}
\end{eqnarray}
The cross sections in Eq.~(\ref{assume}) should be interpreted 
as those for direct production, which means that the feeddown 
contributions from decays of heavier $c \bar c$ states, such as
$\psi(2S) \to \chi_{cJ} \gamma$, have been subtracted.
We will use the assumption that Eq.~(\ref{assume}) 
holds for most production processes to estimate the inclusive 
production rates of $X$ in various processes.  One could 
ignore the motivation based on NRQCD and simply take Eq.~(\ref{assume})
as a phenomenological assumption.

We first consider the inclusive production of $X(3872)$ in $B$ meson 
decays. This decay produces a mixture of $B^+$, $B^0$, and their 
antiparticles. The inclusive branching fraction 
for the direct production of $\chi_{c1}$ from the decays of this mixture 
of $B$ mesons has been measured to be $(3.3 \pm 0.5) \times 10^{-3}$.
(The adjective ``direct'' 
means that the feeddown from the decay $\psi(2S) \to \chi_{c1} \gamma$ 
has been removed.)
The inclusive branching fraction is the sum of the branching fractions 
for the 2-body decay modes $\chi_{c1} K$ and $\chi_{c1} K^*$, 
the 3-body decay modes, etc.  
The fraction of the inclusive decay rate that comes from the 2-body 
decay mode $\chi_{c1} K$ is $(16 \pm 6)$\%.
The fraction of the inclusive decay rate that comes from 2-body decays 
is determined largely by the available hadronic energy, 
which does not vary dramatically among the charmonium states.  
For example, the 2-body decay $B \to J/\psi \, K$ accounts for 
$(11.8 \pm 1.5)$\% of the direct inclusive rate 
for $B \to J/\psi + {\rm anything}$. 
The 2-body decay $B \to \psi(2S) K$ accounts for 
$(18 \pm 4)$\% of the inclusive rate for $B \to \psi(2S) + {\rm anything}$.  
These fractions all differ by less than a standard deviation.
We will use the central value of the observed fraction for $\chi_{c1}$
as an estimate of the fraction for $X$:
\begin{eqnarray}
\frac {{\rm Br} [B \to XK]} {{\rm Br} [B \to X + {\rm anything}]} \approx
\frac {B [B \to \chi_{c1} K]} {B_{\rm dir} [B \to \chi_{c1} + {\rm anything}]} 
\approx 16 \%.
\end{eqnarray}
Combining this with the Belle measurement in Eq.~(\ref{Belle-prod}), 
we obtain the estimate
\begin{eqnarray}
{\rm Br} [B \to X + {\rm anything}] \; {\rm Br} [X \to J/\psi \; \pi^+ \pi^-] 
\approx 8 \times 10^{-5} .
\label{B-pred}
\end{eqnarray}
After multiplying by the 12\% branching fraction of $J/\psi$ into leptons,
the product of the three branching fractions is about $10^{-5}$.
With the Belle and Babar collaborations having each accumulated
more than $10^8$ $B \bar B$ pairs, the predicted rate is large enough 
that the inclusive production of the $X$ should be observable 
in the data from the $B$ factories. 

Dividing the estimate in (\ref{B-pred}) by the 
measured inclusive branching fraction for $B$ decay into $\chi_{c1}$, 
we obtain the product of the constant in (\ref{assume}) 
and the branching fraction for the decay
$X  \to J/\psi \; \pi^+\pi^-$:
\begin{eqnarray}
{\sigma [ X + {\rm anything}] \over \sigma_{\rm dir} [\chi_{c1} + {\rm anything}]} 
 \; {\rm Br} [X \to J/\psi \; \pi^+ \pi^-] 
\approx 0.025.
\label{sigma-pred}
\end{eqnarray}
This can be used to estimate the inclusive production rate of $X$
in any process for which the inclusive production rate of $\chi_{c1}$
has been measured.

In the decay of the $Z^0$, charmonium is produced primarily by the 
decay of the $Z^0$ into $b \bar b$, followed by the fragmentation 
of the $b$ or $\bar b$ into a heavy hadron, 
and then the decay of the heavy hadron into charmonium.
The inclusive branching fraction of $Z^0$ into a charmonium
$H$ is to a good approximation the sum over $b$ hadrons of
the inclusive branching fraction 
for the $b$ hadron to decay into $H$, weighted by the probability 
for a $b$ quark to fragment into the $b$ hadron,
and multiplied by 2 to account for the $b$ and the $\bar b$.
The inclusive branching fraction for the decay of $Z^0$ into $\chi_{c1}$ 
has been measured.  The direct branching fraction obtained by
subtracting the small feeddown from $\psi(2S)$ decays is
$(2.7 \pm 0.8) \times 10^{-3}$.  Multiplying this branching fraction 
by the number in (\ref{sigma-pred}), we obtain the estimate 
\begin{eqnarray}
{\rm Br} [Z^0 \to X + {\rm anything}] \; {\rm Br} [X \to J/\psi \; \pi^+ \pi^-] 
\approx 7 \times 10^{-5} .
\label{Z-pred}
\end{eqnarray}
After multiplying by the 12\% branching fraction of $J/\psi$ into leptons,
the product of the three branching fractions is about $10^{-5}$.
With each of the four detectors at LEP having collected millions of 
hadronic $Z^0$ decays, the predicted rate is large enough that the 
$X$ may be observable in the LEP data.

We next consider the inclusive production of $X(3872)$ in $p \bar p$ 
collisions at the Tevatron.
The inclusive sample of $X$ mesons at the Tevatron consists of two 
components:  a $b$-decay component that comes from inclusive decays of $b$ 
hadrons into $X$, and a prompt component consisting of $X$ mesons 
that are produced directly in the $p \bar p$ collisions.  
The two components can be separated by using 
a vertex detector.  Those $X$ mesons for which the tracks of the pions from 
the decay $X \to J /\psi \; \pi^+ \pi^-$ and the leptons from the decay 
$J/\psi \to \ell^+ \ell^-$ can be traced back to the $p \bar p$ collision 
point are mostly prompt, while those for which the tracks of the charged 
particles can be traced back to a secondary vertex come from $b$ decay.

In Run I of the Tevatron, which consisted of $p \bar p$ collisions 
at center-of-mass energy of 1.8 TeV,
the CDF collaboration measured the fraction of prompt $J/\psi$'s 
that come from $\chi_c$ decay and the ratio of the prompt 
cross sections for producing $\chi_{c1}$ or $\chi_{c2}$.
The fraction of prompt $J/\psi$'s with transverse momentum $p_T > 4$ GeV
and pseudorapidity satisfying $|\eta| < 0.6$
that come from the decays of $\chi_{c1}$ or $\chi_{c2}$ is
$0.297 \pm 0.017({\rm stat}) \pm 0.057({\rm syst})$ \cite{Abe:1997yz}.
The ratio of the prompt cross sections for producing
$\chi_{c1}$ and $\chi_{c2}$, where the $\chi_c$ subsequently decays into 
$J/\psi \; \gamma$ with the $J/\psi$ having $p_T < 4$ GeV,
is $1.04 \pm 0.29({\rm stat}) \pm 0.12({\rm syst})$ \cite{Affolder:2001ij}.
Combining these results with the estimate in (\ref{sigma-pred}),
we obtain the estimate
\begin{eqnarray}
{\sigma [ X + {\rm anything}] \over \sigma [ J/\psi + {\rm anything}] }
 \; {\rm Br} [X \to J/\psi \; \pi^+ \pi^-] 
\approx 4 \times 10^{-3}.
\label{ppbar-pred}
\end{eqnarray}
The left side of Eq.~(\ref{ppbar-pred}) is simply
the fraction of prompt $J/\psi$'s that come from decays of $X$
multiplied by the ratio of the branching fractions for 
$X \to J/\psi \; \pi^+ \pi^-$ and $X \to J/\psi \; +{\rm anything}$.
The fraction of prompt $J/\psi$'s that come from decays of $\psi(2S)$
are observed to increase with $p_T$ from $(7\pm 2)$\% at $p_T=5$ GeV
to $(15\pm 5)$\% at $p_T=18$ GeV \cite{Abe:1997yz}. 
In contrast, the fraction of $J/\psi$'s that come from decays of 
$\chi_c$ seemed to decrease slightly with $p_T$ \cite{Abe:1997yz}.
In Run II of the Tevatron, improvements in the detectors will allow 
these fractions to be measured as a function of $p_T$ down to $p_T=0$.
Our assumption (\ref{assume}) implies that the 
fraction of prompt $J/\psi$'s that come from decays of $X$ 
should have the same $p_T$-dependence as the fraction of $J/\psi$'s 
that come from decays of $\chi_{cJ}$. 

We have pointed out that the NRQCD factorization formalism that 
describes inclusive charmonium production can also be applied 
to the inclusive production of pairs of charmed mesons
with sufficiently small invariant mass.  If the $X(3872)$ is a 
$D^{*0} \bar D^0$/$D^0 \bar D^{*0}$ molecule, the NRQCD factorization 
formalism also applies to the inclusive production rate of $X$.
We argued that if $X$ has quantum numbers $1^{++}$,
the leading term in the NRQCD factorization formula for $X$
should be the color-octet $^3S_1$ term.
We used the assumption that the ratio of the 
inclusive production rates of $X$ and $\chi_{cJ}$ 
should be roughly the same for all processes
to estimate the inclusive production rate of 
$X(3872)$ in various high energy processes.
Measurements of inclusive production rates consistent with these 
estimates would be consistent with the interpretation
of the $X(3872)$ as 
a $D^{*0} \bar D^0$/$D^0 \bar D^{*0}$ molecule with $J^{PC} = 1^{++}$. 

I thank G.T.~Bodwin for a useful discussion.
This research was supported in part by the Department of Energy under
grant DE-FG02-91-ER4069.


\end{document}